\documentclass[%
 aip,
 amsmath,amssymb,
 reprint,%
]{revtex4-1}
\usepackage{amssymb}
\usepackage{tcolorbox}
\usepackage{graphicx}
\usepackage{dcolumn}
\usepackage{bm}

\usepackage[utf8]{inputenc}
\usepackage[T1]{fontenc}
\usepackage{mathptmx}
\usepackage{etoolbox}
\usepackage{enumerate}
\usepackage{engord}
\usepackage{placeins}
\usepackage{booktabs}
\usepackage{threeparttable}
\usepackage{multirow}

\makeatletter
\def\@email#1#2{%
 \endgroup
 \patchcmd{\titleblock@produce}
  {\frontmatter@RRAPformat}
  {\frontmatter@RRAPformat{\produce@RRAP{*#1\href{mailto:#2}{#2}}}\frontmatter@RRAPformat}
  {}{}
}%
\makeatother
\begin{document}

\preprint{AIP/123-QED}

\title{Physics-informed machine learning of the correlation functions in bulk fluids}
\author{Wenqian Chen}
\altaffiliation{Advanced Computing, Mathematics and Data Division, Pacific Northwest National Laboratory, Richland, WA, 99354, USA}
\author{Peiyuan Gao}
\email{peiyuan.gao@pnnl.gov}
\altaffiliation{Advanced Computing, Mathematics and Data Division, Pacific Northwest National Laboratory, Richland, WA, 99354, USA}

\author{Panos Stinis}
\altaffiliation{Advanced Computing, Mathematics and Data Division, Pacific Northwest National Laboratory, Richland, WA, 99354, USA}

\begin{abstract}
The Ornstein-Zernike (OZ) equation is the fundamental equation for pair correlation function computations in the modern integral equation theory for liquids. In this work, machine learning models, notably physics-informed neural networks and physics-informed neural operator networks, are explored to solve the OZ equation. The physics-informed machine learning models demonstrate great accuracy and high efficiency in solving the forward and inverse OZ problems of various bulk fluids. The results highlight the significant potential of physics-informed machine learning for applications in thermodynamic state theory.
\end{abstract}

\maketitle

\section{\label{sec:level1}Introduction}
Statistical mechanics has been successful as a framework for modeling complex  systems from microscale to macroscale.\cite{Kardar2007,Tuckerman2010}
The integral equation theory (IET) approach for liquids is one of the most widely used approaches that is based on statistical mechanics.\cite{Hirata2003, Hansen2013,cao2019}
It provides a closed analytical relation between the molecular interaction potentials and microscopic correlation functions of liquids and liquid mixtures.
The prediction of macroscopic properties from the knowledge of the microscopic structure allows for a detailed description of a wide diversity of bulk fluids.
The basic concept in the theory is the Ornstein–Zernike (OZ) equation.\cite{Ornstein1914}
With the aid of statistical mechanics, along with some approximation methods such as hypernetted chain (HNC), Percus-Yevick (PY), and Verlet modified (VM)\cite{Hansen2013, Choudhury2002,Tsednee2019,CARVALHO20212,Bedolla2022} closure approximations, 
many properties of bulk fluids can be calculated. 
The OZ equation has practical importance as a foundation for approximations for computing the pair correlation function of molecules or ions, or of colloidal particles in solution.\cite{Eckert2020,Tschopp2021}
Furthermore, based on the OZ equation, several approaches such as the molecular Ornstein–Zernike (MOZ) equation\cite{Richardi1998} and the three-dimensional reference interaction site model (3D-RISM) theory\cite{beglov1997,Yoshida2017,cao2023}
have been developed. These theories  offer a rigorous framework for calculating equilibrium solvation properties  without the need for costly dynamic simulations, which can be crucial for biochemical process and drug design.\cite{Sindhikara2013,Samways2021,Ratkova2015} The integral equation theory of molecular liquids has been an active area of academic research. 

In recent years, the development of machine learning (ML) techniques offers a suite of powerful tools in function approximation and equation solving.\cite{goodfellow2016deep} ML approaches have been successfully applied to problems such as solving the direct correlation function of materials\cite{carvalho2020radial,Carvalho2022} or learning  closures within the OZ framework\cite{goodall2021data}.
Among ML techniques, physics-informed neural networks\cite{raissi2019physics} (PINNs) ,  are a novel type of neural network that exploit the known laws of physics by including them in the loss function. PINNs are designed to address systems governed by established laws of physics. Their versatility allows them to tackle a range of challenges, ranging from elementary algebraic equations to sophisticated physical phenomena, notably fluid dynamics\cite{cai2021physicsfluid, raissi2020hidden, chen2021physics}, heat transfer\cite{cai2021physics, hennigh2021nvidia}. However, PINNs can sometimes struggle with the complexities introduced by parameterized partial differential equations (PDEs). Such PDEs can include parameters that are related to specific material properties, which can  increase significantly the required complexity of a general-purpose  neural network. The deep operator network\cite{lu2021learning} (DeepOnet)  is designed for operator learning, where the inputs are processed separately and then combined to produce the final output. Hence, by substituting the general-purpose neural network with DeepOnet within the PINN framework, the physics-informed deep operator network (PIDeepOnet) is well-suited for handling parameterized PDEs.
Currently, the application of PINN or PIDeepOnet in IET is relatively rare. Carvalho and Braga \cite{carvalho2022physics} developed a PINN model to solve the OZ equation and obtained impressive prediction results for the direct correlation function, showcasing the potential of PINNs in solving OZ equations. However, the complexity of their neural network architecture results in slow training. Furthermore, there is potential for enhancing prediction accuracy, particularly for the long-range correlation at higher densities. 

In this work, we build a new physics-informed machine learning framework. This includes a PINN approach tailored for {\it fixed-parameter} OZ equations and a PIDeepOnet approach for {\it parameterized} OZ equations. By integrating IET with cutting-edge machine learning techniques—like Fourier feature embedding\cite{tancik2020fourier}, the modified feedforward neural network\cite{wang2021understanding}, and the self-adaptive weighting method\cite{mcclenny2020self}, we achieve marked improvements in accuracy and computational efficiency compared to the existing PINN approach.\cite{carvalho2022physics}. Furthermore, the PIDeepOnet offers the potential to apply IET for fluids across diverse thermodynamic states \cite{Jaiswal2014, Chandler1973}.

\section{\label{sec_method}Methodology}
\subsection{\label{sec_OZ}The OZ equation approach to the physics of simple fluids}
As a rigorous framework in statistical mechanics, the OZ equation \cite{ornstein1914influence} plays a vital role in approximating  correlations among particles in fluids. The OZ equation has practical importance as a foundation for approximations for computing the pair correlation function of molecules or particles in liquids. The pair correlation function is related via Fourier transform to the static structure factor, which can be determined experimentally using X-ray diffraction or neutron diffraction. For the simplest case, i.e., a homogeneous isotropic fluid with density $\rho$, the total correlation function $h(r)$ is given by
\begin{equation}
	h(r)=g(r)-1,
\end{equation}
where the function $g(r)$ is the two-body correlation function. It is the probability of finding a particle a distance $r$ from another particle at 0. This probability in a liquid becomes one for large enough $r$. 
The OZ equation establishes a relationship between 
the total correlation function $h(r)$ and the direct correlation function $c(r)$ as follows:
\begin{equation}
	h(r)=c(r)+\rho\int{ c(|r-r'|) h(r')dr' },
\end{equation}
where $\rho$ is the number density of the system.
As both functions $h(r)$ and $c(r)$ are unknown, an additional equation is required, which is known as a closure relation. This closure typically involves a physically-motivated approximation, which greatly impacts the resultant solution. The commonly accepted form for the closure is
\begin{equation}
	h(r)+1=\exp(-\beta u(r) + \gamma(r) + B(r)),
\end{equation}
where  $\gamma(r)=h(r)-c(r)$ is the indirect correlation function, $\beta=1/(k_B T)$, $k_B$ is Boltzmann constant and $T$ is the temperature. $B(r)$ is the bridge function defined for specific applications and it can be expressed as a power series in $\rho$ of irreducible diagrams. $u(r)$ represents the interaction potential between particles for a specific fluid. In this work, we opt to use the 12-6 Lennard-Jones (LJ) potential  unless stated otherwise, which is defined as
\begin{equation}
	u(r)=4\epsilon\left[ 
	\left(\frac{\sigma}{r}\right)^{12}-
	\left(\frac{\sigma}{r}\right)^{6}
	 \right],
\end{equation}
where $\epsilon$ and $\sigma$ are the energy and size parameters, respectively.
 In this work, we tested three classical closures, namely the  HNC closure \cite{Hansen2013},  PY closure \cite{Hansen2013}  and VM closure \cite{Choudhury2002}. 

The PY closure, where $B(r)=\ln(1+\gamma(r))-\gamma(r)$ uses a linear approximation effective for weakly correlated systems, especially hard-sphere systems. In contrast, the HNC closure where $B(r)=0$ employs a non-linear approximation that better describes strongly correlated systems, although with more computational complexity. The VM closure offers a balance between the PY and HNC closures, providing an alternative for hard-sphere systems. The bridge function for the VM closure is defined as follows:
\begin{equation}
	B(r)=\left\{
		\begin{aligned}
	    &\frac{-s^2}{2(1+\alpha s)}  &\text{for} \quad &s\ge 0 \\
	    &-\frac{1}{2}s^2             &\text{for} \quad &s<0
	    \end{aligned}
	    \right .
	    ,
\end{equation}
where $\alpha=1.01752-0.275\rho^*$, $\rho^*=\rho\sigma^3$   and $s(r)=\gamma(r)-\beta u_2(r)$. The function $u_2(r)$ is the perturbation potential \cite{Choudhury2002, duh1996integral}, which is defined as 
\begin{equation}
	u_2(r)=-4\epsilon \left(\frac{\sigma}{r}\right)^6\exp\left(-\frac{1}{\rho^*}\left(\frac{\sigma}{r}\right)^{6\rho^*}\right)
	.
\end{equation}
\subsection{\label{sec:_PIML}Physics-informed machine learning of OZ equation}

In order to solve the OZ equation by physics-informed machine learning techniques, a neural network is employed to approximate the solution  with respect to the input (the network architecture will be introduced in subsection \ref{sec_networks}). In general, the network mapping is defined as $\mathbf{y}(\mathbf{x}; \pmb{\theta})$, where $\mathbf{x}$ and $\mathbf{y}$ are the input and output of the network, respectively, and $\pmb{\theta}$ represents the trainable parameters (weights and biases) of the network. Physical laws consist of governing equations, often formulated as partial differential equations (PDEs), and their respective boundary conditions. In general, the loss function, when constrained by physical laws, can be represented in a composite manner, namely
\begin{equation}
	\label{eq_loss}
	\begin{aligned}
	\mathcal{L}(\pmb{\theta}) & = \mathcal{L}_{PDE}(\pmb{\theta}) + \mathcal{L}_{BC}(\pmb{\theta}) \\
	     & = \frac{1}{N_r}\sum_{i=0}^{N_r}{\mathcal{N}^2(\mathbf{y}(\mathbf{x}_i^r))} + \frac{1}{N_b}\sum_{i=0}^{N_b}{\mathcal{B}^2(\mathbf{y}(\mathbf{x}_i^b))},
	\end{aligned}
\end{equation}
where $\mathcal{N}$ and $\mathcal{B}$ represent the residuals of the operator corresponding to the governing equations and the boundary conditions, respectively. $N_r$ and $N_b$ are the sizes of residual data set $\{\mathbf{x}_i^r\}_{i=1}^{N_r}$ and boundary data set $\{\mathbf{x}_i^b\}_{i=1}^{N_b}$.

The network is trained by minimizing the loss function. To cope with the complexity engendered by the multiple components of the loss function, 
it is  essential to further balance these terms with adaptive weights \cite{wang2022and}. During training, we adopt a recently proposed self-adaptive weighting method   \cite{mcclenny2020self}, which is briefly introduced in the following. By assigning a weight to each individual training point, the loss function in Eq. \eqref{eq_loss} is recast as 
\begin{equation}
	\label{eq_loss_SA}
	\begin{aligned}
	\mathcal{L}(\pmb{\theta}, \mathbf{w})&=
	 \frac{1}{N_r}\sum_{i=0}^{N_r}{ M( w_i^r)\mathcal{N}^2(\mathbf{y}(\mathbf{x}_i^r))} \\
	 &+ \frac{1}{N_b}\sum_{i=0}^{N_b}{M( w_i^b) \mathcal{B}^2(\mathbf{y}(\mathbf{x}_i^b))}		
	 \end{aligned}
 ,
\end{equation}
where $\mathbf{w}$ represents the collection of the trainable weights and $M$ denotes a mask function. In this study, we adopt the mask function $M(x) = x^2$, a choice that aligns with the approach taken in reference\cite{mcclenny2020self}. The optimization of the loss function given by Eq. \eqref{eq_loss_SA}, involves minimization with respect to the parameters $\pmb{\theta}$, while concurrently maximizing with respect to the weights $\mathbf{w}$, namely
\begin{equation}
	\min_{\pmb{\theta}} \max_{\mathbf{w}}  \mathcal{L}(\pmb{\theta}, \mathbf{w}).
\end{equation}
In the context of a gradient ascent/descent approach, the  updating of parameters and weights is carried out simultaneously, namely
\begin{equation}
	\begin{aligned}
		\pmb{\theta}^{k+1}  &= \pmb{\theta}^{k} - \eta_{\pmb{\theta}}^k \nabla_{\pmb{\theta}}\mathcal{L}(\pmb{\theta}^k, \mathbf{w}^k) \\
		\mathbf{w}^{k+1} &= \mathbf{w}^{k} + \eta_{\mathbf{w}}^k \nabla_{\mathbf{w}}\mathcal{L}(\pmb{\theta}^k, \mathbf{w}^k) \\
	\end{aligned} 
\end{equation}
where $k$ represents the training iteration index, while $\eta_{\pmb{\theta}}^k$ and $\eta_{\mathbf{w}}^k$ are the learning rates for the parameters and self-adaptive weights, respectively. 
The gradient of the loss function with respect to the weights is computed as follows
\begin{equation}	
	\begin{aligned}
		\nabla_{\mathbf{w}}\mathcal{L}(\pmb{\theta}, \mathbf{w})
		&=
		\left\{
		M^{\prime}(w_i^r)\mathcal{N}^2(\mathbf{x}_i^r) 
		\right\}_{i=1}^{N_{r}}             \\
		&\cup
		\left\{
		M^{\prime}(w_i^d)\mathcal{B}^2(\mathbf{x}_i^b)	\right\}_{i=1}^{N_{b}},      
	\end{aligned} 	\label{eq_gradien_SA2}
\end{equation}
where $M'$ is the derivative of the mask function $M$. It is worth noting that Eq. \eqref{eq_gradien_SA2} can be calculated manually instead through automatic differentiation to save computational cost.
In this work, we employ the OZ equation as well as the specific closure to build the physics-informed loss function to supervise the training of the network. In this work, the OZ equation (an integral equation) is occasionally termed as a PDE. Readers are advised to recognize this terminological choice.

\subsection{Network architecture}
\label{sec_networks}
For the OZ equation with fixed parameters, we use a modified feedforward neural network (FNN) recently introduced in the reference\cite{wang2021understanding}, which has demonstrated to be more effective than the standard FNN in mapping between input and output.

A modified FNN maps the input $\mathbf{x}$ to the output $\mathbf{y}$. Generally, a modified FNN consists of an input layer, $L$ hidden layers and an output layer.  The $l$-th layer has $n_l$ neurons, where $l=0,1,..L,L+1$ denotes the input layer, first hidden layer,..., $L$-th hidden layer and the output layer, respectively. Note that the number of neurons of each hidden layer is the same, i.e., $n_1=n_2=...=n_L$. The forward propagation, i.e., the function $\mathbf{y}=f_{\text{nn}}(\mathbf{x})$  is
\begin{equation}
	\left\{
	\begin{aligned}
		\mathbf{U}\;&= f_{act}(\mathbf{W}^U\mathbf{x}+\mathbf{b}^U) &&\\
		\mathbf{V}\;&= f_{act}(\mathbf{W}^V\mathbf{x}+\mathbf{b}^V) &&\\
		\mathbf{y}^{1} &= f_{act}( \mathbf{W}^{1}\mathbf{x}+ \mathbf{b}^{1}) &&\\
		\mathbf{Z}^{l} &= f_{act}( \mathbf{W}^{l}\mathbf{y}^{l-1}+ \mathbf{b}^{l}), && 2\le l \le L \\
		\mathbf{y}^{l} &= (1-\mathbf{Z}^{l}) \odot \mathbf{U} + \mathbf{Z}^{l} \odot \mathbf{V}, && 2\le l \le L \\
		\mathbf{y} \;     &=\mathbf{y}^{L+1}=\mathbf{W}^{L+1}\mathbf{y}^{L}+ \mathbf{b}^{L+1} &&\\
	\end{aligned}
	\right .
	,
\end{equation}
where $\mathbf{x} \in \mathbb{R}^{n_0}$ is the input, $\odot$ denotes point-wise multiplication, $\mathbf{y}^{l}  \in \mathbb{R}^{n_{l}} $ is the output of the $l_\text{th}$ layer, $\mathbf{W}^{l} \in \mathbb{R}^{n_{l}} \times \mathbb{R}^{n_{l-1}}$ , $\mathbf{W}^U \in \mathbb{R}^{n_1} \times \mathbb{R}^{n_0} $ and $ \mathbf{W}^V \in \mathbb{R}^{n_1} \times \mathbb{R}^{n_0}$ are the weights, and
$\mathbf{b}^{l} \in \mathbb{R}^{n_{l}} $, $\mathbf{b}^{U} \in \mathbb{R}^{n_1}$ and  $\mathbf{b}^{V} \in \mathbb{R}^{n_1}$ are the biases.  $f_{act}$ is a point-wise activation function, and we chose the Swish activation function $f_{act}(x)=x\cdot \text{sigmoid}(x)$ \cite{ramachandran2017searching}.

For solving parameterized PDEs, the  parameter is also taken as a component of the input. We resort to DeepOnet to build the input-output mapping \cite{lu2021learning}. A DeepOnet is made of two sub-networks, the trunk and branch, where the truck network takes spatial coordinates as input and the branch network takes the parameter as input. The output of  DeepOnet is a function of the truck and branch outputs, which reads:
\begin{equation}
	y=\sum_{i=1}^{I}{b_i t_i}+b_0,
\end{equation}
where $\mathbf{b}=\{b_i\}_{i=1}^{I}$ and $\mathbf{t}=\{t_i\}_{i=1}^{I}$ are the branch  and trunk outputs, respectively. $b_0$ is a trainable bias variable. As for the architecture of the trunk and branch networks, we also adopt the modified FNN.
So the forward propagation for the DeepOnet, i.e., the function   $y=f_{\text{DeepOnet}}(\mathbf{x})$ is:
\begin{equation}
	\left\{
	\begin{aligned}
	\mathbf{t}&=f_{\text{nn,t}}(\mathbf{x}_t)\\
	\mathbf{b}&=f_{\text{nn,b}}(\mathbf{x}_b)\\
	y&=\sum_{i=1}^{M}{b_i t_i+b_0}
	\end{aligned}
	\right .
	.
\end{equation}
where $f_{\text{nn,t}}$ and $f_{\text{nn,b}}$ are the forward propagation functions of the trunk and branch nets, respectively. Also, $\mathbf{x}_t$ and $\mathbf{x}_b$ are the trunk and branch inputs, respectively.

\subsection{\label{sec:level2}Data preprocessing}
The effectiveness of neural network training is often influenced by the scale of the input/output data, as highlighted by the reference \cite{loffe2014accelerating}. Consequently, implementing normalization prior to training can significantly improve performance.
In a given problem scenario, the lower and upper boundaries of the input vector $\mathbf{x}$ are often predefined. These can be employed to scale the input $\mathbf{x}$, making it lie within the interval $[-1, 1]$ for each dimension:
\begin{equation}
	\begin{aligned}
		\widetilde{\mathbf{x}} &= s_{I}\left( \mathbf{x} \right)
		= \frac{\mathbf{x}-(\mathbf{x}_{\max}+\mathbf{x}_{\min})/2}{(\mathbf{x}_{\max} -\mathbf{x}_{\min})/2}		
	\end{aligned}
	\label{eq_norm_input}
\end{equation}

In our case, the indirect correlation function $\gamma(r)$ is known to oscillate around 0 with waves whose spectrum covers a wide range. However, PINN training can suffer from a spectral bias problem \cite{basri2020frequency,rahaman2019spectral,wang2021eigenvector}, namely, the network tends to focus on learning to low-frequency modes. To address this issue in the context of the OZ equation, we apply a straightforward transformation of the normalized coordinate $\widetilde{r}$ into multiple Fourier features, as introduced in the reference\cite{tancik2020fourier}. The transformation is constructed by a random Fourier mapping $s_F: \widetilde{r} \to \mathbb{R}^{2m}$:
\begin{equation}
	\label{eq_featureTransform}
	s_F(\widetilde{r}) =
	\left( 
	\sin(\pi \mathbf{B} \widetilde{r}),
	\cos(\pi \mathbf{B} \widetilde{r})
	\right), 
\end{equation}
where $\mathbf{B} \in \mathbb{R}^{m}$ is a random vector and $m$ is the number of Fourier features.  The elements of $\mathbf{B}$ are sampled from a Gaussian distribution $b\sim N(0,\sigma_F ^2)$, where $\sigma_F^2$ is the feature variance.
Embedding Fourier features can enhance the network's ability to learn high-frequency modes. Using Fourier feature embedding with increased variance generally aids in capturing higher-frequency modes, though it may necessitate more residual points. 
Consequently, it's essential to select the Fourier feature variance carefully based on the specific problem at hand.
For more details about Fourier feature embedding, we refer to the reference \cite{tancik2020fourier}.

For the output, a predefined range is not always available. Since there is usually not any prior knowledge about the output, the normalization is just the identical mapping, namely
\begin{equation}
	\widetilde{\mathbf{y}} = s_{O}\left( \mathbf{y} \right) =\mathbf{y}.
	\label{eq_norm_output1}	
\end{equation}

\subsection{Training details }
We initiate the network using the Xavier initialization method \cite{glorot2010understanding}, with weights initialized accordingly and biases set to zero. Initially, the self-adaptive weights are assigned a value of 1. The network training undergoes a two-stage process: In the first stage, we employ the Adam optimizer \cite{kingma2014adam} for 36,000 iterations. The learning rates are fixed as $\eta_{\pmb{\theta}} =0.0001$ and $\eta_{\mathbf{w}} = 0.1$. In the subsequent stage, the next 4,000 iterations use the L-BFGS optimizer \cite{liu1989limited}, during which the self-adaptive weights remain unchanged. 

Our training approach is implemented in PyTorch \cite{paszke2019pytorch} and is executed on a GPU cluster. For computations, we use a 32-bit single-precision data type and a single NVIDIA\textsuperscript{\textregistered} Tesla P100 GPU.

We observed that during the L-BFGS updating phase, the training results can suddenly explode, increasing to very large magnitudes. This behavior might be attributed to the complex nonlinearity nature of the OZ equation. To counteract this, we have selected the best-trained network from prior to this explosion as our final model.

\section{\label{sec_result}Result and discussion}

\subsection{Solution of OZ equation for simple fluid}
\label{sec_PINN_forward_inverse}

\subsubsection{Forward problem} \label{sec_PINN_forward}
We first test the capability of the PINN approach for the OZ equation with the HNC closure. The reason we choose the HNC closure is that a previous PINN model\cite{carvalho2022physics} struggled with predicting long-range correlations due to its inherent nonlinearity.
The parameters in the OZ equation are set as $k_B=\sigma=\epsilon=1$ and $T=2$. The unknown of the OZ equation, namely the indirect correlation $\gamma(r)$ is characterized by a sharp gradient for $r \to 0$, which presents a significant challenge for the training of networks. 
In line with the strategy presented by \cite{carvalho2022physics}, our approach takes $r$ as input, and predicts $\Gamma(r)=r\gamma(r)$ as opposed to $\gamma(r)$. This strategy assists in moderating the sharpness of the gradient, thereby facilitating improved training of the network.
Similarly, we define $C(r)=rc(r)$ for the direct correlation, whose Fourier transformation \cite{lado1971numerical} is calculated as 
\begin{equation}\label{eq_C_Fourier}
	\widehat{C}(q_j)=4\pi \Delta r \sum_{i=1}^{N-1}{C(r_i)}\sin(r_i q_j)
 ,
\end{equation} 
where $r_i=i\Delta r$ for $i=0,1,...,N$, $q_j=j\Delta q$ for $j=0,1,...,N$, $\Delta r=\frac{R_{\max}}{N}$ and $\Delta q=\frac{\pi}{R_{\max}}$. Here, $R_{\max}=20$ is the maximum $r$ coordinate to be studied and $N=800$ is chosen.
In the reciprocal space for the OZ equation, we obtain the Fourier transformation of $\Gamma(r)$  according to  $\widehat{C}(q)$, namely
\begin{equation}\label{eq_Gamma_OZ}
	\widehat{\Gamma}(q_j)=\frac{\rho  \widehat{C}^2(q_j)}
	{q_j-\rho \widehat{C}(q_j)}
 .
\end{equation}
Then an inverse Fourier transformation is performed to recover $\Gamma(r)$, namely
\begin{equation}\label{eq_Gamma_inverse_Fourier}
	\Gamma^*(r_i)=\frac{\Delta q}{2\pi ^2}
	\sum_{j=1}^{N-1}{\widehat{\Gamma}(q_j)\sin(r_i q_j})
 .
\end{equation}
$\Gamma(r)$ is the prediction of network and $\Gamma^*(r)$ is the prediction encoded by the physical laws of OZ equation. When they are equal, the OZ equation is solved. Thus, the PDE component of the loss function  is built as the mean squared error between them, namely
\begin{equation}\label{eq_loss_PDE}
	\mathcal{L}_{PDE}=\frac{1}{N-1}\sum_{i=1}^{N-1}\left[{\Gamma^*(r_i)-\Gamma(r_i)}\right]^2
\end{equation}
The boundary condition is defined as 
\begin{equation}
	\Gamma(0)=\Gamma(R_{\max})=0.
\end{equation}
Thus the loss function is defined as
\begin{align}
	loss&=\mathcal{L}_{PDE}+\mathcal{L}_{BC}\\
	    &=\frac{1}{N-1}\sum_{i=1}^{N-1}\left[{\Gamma^*(r_i)-\Gamma(r_i)}\right]^2 + \frac{1}{2}\left[\Gamma^2(0)+\Gamma^2(R_{\max})\right]
	    .
\end{align}

\begin{figure*}[htbp]
	\centering	
	\includegraphics[width=14cm]{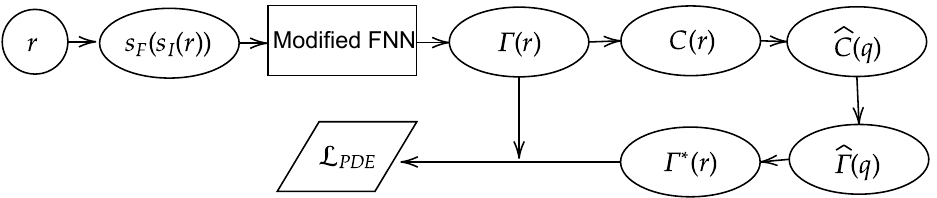}
	\caption{Forward propagation of PDE loss in the PINN approach for solving the OZ equation. $s_I(\cdot)$ denote the normalization function of input and $s_F(\cdot)$ denotes the Fourier feature transformation function.}
	\label{fig_PINN_forward}	
\end{figure*}

For the sake of clarity, the forward propagation for deriving the PDE loss is shown in 
Fig. \ref{fig_PINN_forward}. The steps for calculating the PDE loss are listed as follows:
\begin{enumerate}[(1)]
\item Predict  $\Gamma(r)$ at $N-1$ equally-spaced points $r=r_i$ for $i=1,...,N-1$.
\item Compute $C(r)$ according to the HNC closure.
\item Perform the Fourier transformation of $C(r)$ according to Eq. \eqref{eq_C_Fourier}.
\item Calculate the Fourier transformation of $\Gamma(r)$ with Eq. \eqref{eq_Gamma_OZ}.
\item Perform the inverse Fourier transformation in Eq. \eqref{eq_Gamma_inverse_Fourier} to recover $\Gamma^*(r).$
\item Calculate the PDE loss with Eq. \eqref{eq_loss_PDE}.
\end{enumerate}

\begin{figure*}[htbp]
	\centering	
	\includegraphics[width=14cm]{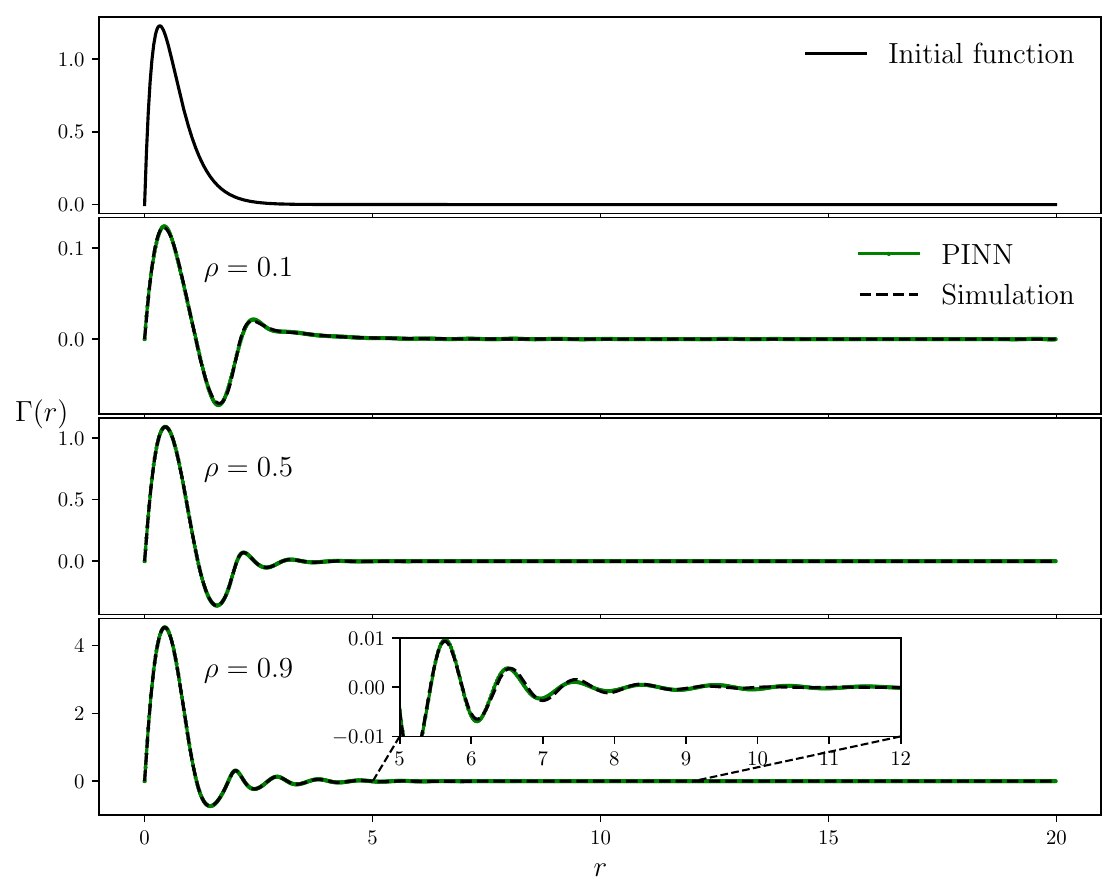}
	\caption{Initial function (\engordnumber{1} row) and PINN predictions of $\Gamma(r)$ vs. simulation results for the OZ equation.}
	\label{fig_PINN_forward_result}	
\end{figure*}

The network is structured as  $(1, 30, 30, 30, 30, 30, 30, 1)$, with each entry denoting the neuron count for that layer. The feature variance $\sigma_F^2=1$ and the feature number $m=50$ are set for the Fourier feature embedding.
As pointed out in reference \cite{carvalho2022physics}, proper initialization of a network greatly contributes to efficient and effective network training. For this purpose, we first train the network with an initialization function $y=cre^{-br}$ with $c=10$ and $b=3$. 
The function applies for all densities in this section and is presented in Fig. \ref{fig_PINN_forward_result}. Then we train the network with the PINN approach starting from the pre-trained network. 
Fig. \ref{fig_PINN_forward_result} showcases a comparison between PINN predictions and numerical results. The latter, deemed as the ground truth, is obtained via the iterative Fourier method at a high resolution of 
$N=2000$.
The PINN approach is able to predict the solutions accurately with a relative $L_2$ error $2.04\times 10^{-3}, 4.11 \times 10^{-4}, 2.40 \times 10^{-4}$ for density $\rho=0.1, 0.5, 0.9$, respectively. The performance of our model is similar to a previous PINN model\cite{carvalho2022physics} when the density of the simulation system is not high, e.g., less than 0.5. With the increase of density, the solution is shown to oscillate with higher frequency and larger amplitude due to the long-range correlation. Even for the large density case $\rho=0.9$, the prediction of our PINN model can still accurately capture the oscillation of small amplitude (about 0.01), as shown in the close-up view in Fig. \ref{fig_PINN_forward_result}. The superiority of the present PINN approach is evident, compared with results in the reference \cite{carvalho2022physics} which fails to capture the long-range correlation ($5\le r\le 8$). This indicates that our model has a wider scope of application for various fluids. In addition, the training time is less than 10 minutes with a single Nvidia P100 graphics card, while the model in the previous paper would take several hours to train \cite{carvalho2022physics}. This could be attributed to the complex neural network architecture of their PINN model.  

\subsubsection{Inverse problem}
The two-body static correlation function $g(r)$ is the quantity of interest in radiation scattering experiments. In this section,  we use PINN to infer $g(r)$ from experimental neutron scattering data $s(q)$ (structure factor). Their relation is reflected in the following equation:
\begin{equation}\label{eq_inverse}
	s(q) = 1+4\pi \rho \int
	{\left[g(r)-1\right]r^2\left[\frac{sin(qr)}{qr}\right]}.
\end{equation}
We use  experimental data for liquid argon at 85 K.\cite{Yarnell1973} The reduced number density is $\rho=1000/36.0385^3$. The maximum  $r$ coordinate is chosen as $R_{\max}=36$.

The network takes $r$ as input and outputs $g(r)$. 
The loss function is built as the mean squared error between the prediction of $s(q)$ using the network for $g(r)$ in Eq. \eqref{eq_inverse} and its experimental observation $s_{exp}(q)$, namely
\begin{equation}\label{eq_loss_inverse}
	\mathcal{L} = \frac{1}{N_q}\sum_{i=1}^{N_q}\left[s(q_i)-s_{exp}(q_i)\right].
\end{equation}
where the experimental data of $s_{exp}(q)$ contains $N_q=400$ points within the range $q\in (0, 11.8)$. To evaluate $s(q)$ using the network for $g(r)$, the integral in Eq. \eqref{eq_inverse} is approximated by the compound trapezoidal formula with $N_r+1$ equally-spaced points over the range $[0, R_{\max}]$. $N_r$ is set as 800.

\begin{figure*}[htbp]
	\centering	
	\includegraphics[width=14cm]{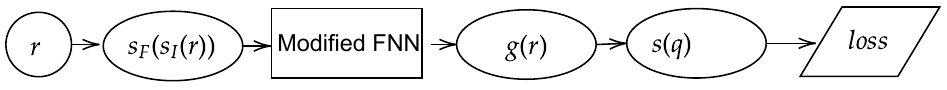}
	\caption{Forward propagation in the PINN approach for solving the inverse problem. $s_I(\cdot)$ denote the normalization function of input and $s_F(\cdot)$ denotes the Fourier feature transformation function.}
	\label{fig_PINN_inverse}	
\end{figure*}

For the sake of clarity, the forward propagation for the PINN approach is shown in Fig. \ref{fig_PINN_inverse}. The steps for calculating the loss are listed as follows:
\begin{enumerate}[(1)]
	\item Predict  $g(r)$ at $N_r+1$ equally-spaced points
	\item Compute $s(q_i)$ for $i=1,...,N_q$ according to Eq. \eqref{eq_inverse}.
	\item Calculate the loss with Eq. \eqref{eq_loss_inverse}.
\end{enumerate}

\begin{figure*}[htbp]
	\centering	
	\includegraphics[width=15.5cm]{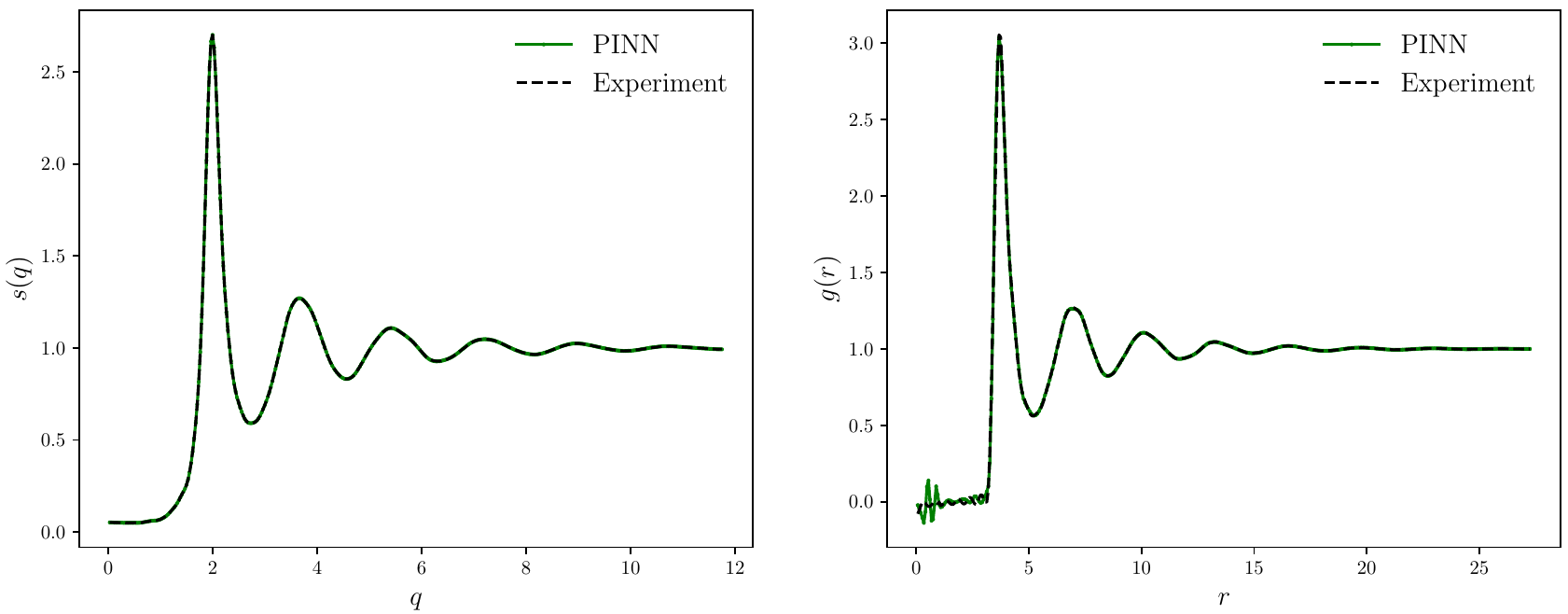}
	\caption{PINN predictions of $s(q)$ and $g(r)$ vs. experimental observations for the inverse inference of radial distribution $g(r)$.}
	\label{fig_PINN_inverse_result}	
\end{figure*}

The network is structured as  $(1, 30, 30, 30, 30, 30, 30, 1)$, with each entry denoting the neuron count for that layer. The feature variance $\sigma_F^2=1$ and the feature number $m=50$ are set for the Fourier feature embedding.
Similarly, the network is pre-trained with a initialization function, which is defined as the low density limit of the radial distribution, namely $g(r)=e^{-\frac{u(r)}{k_BT}}$ with $\epsilon=\sigma=1$, $k_B=1$ and $T=2$. The PINN predictions are shown in Fig. \ref{fig_PINN_inverse_result}. The PINN prediction of $s(q)$ coincides well with the experimental data with a relative $L_2$ error $5.14 \times 10^{-4}$, implying a tiny training error. In the region where $r$ is small, the PINN prediction of $g(r)$ exhibits spurious fluctuations. This behavior can be neglected because the exact $g(r)$ in this region should be 0, which is also noted in the references \cite{carvalho2022physics,carvalho2020neural}. For a larger $r$, the PINN prediction of $g(r)$ agrees well with the experimental data, and the relative $L_2$ error is $4.48 \times 10^{-3}$,  when calculated over the region $3.5<r<R_{\max}$. 

\subsection{Comparison between various closures}
\label{sec_closures}
\begin{figure*}[htbp]
	\centering	
	\includegraphics[width=15.5cm]{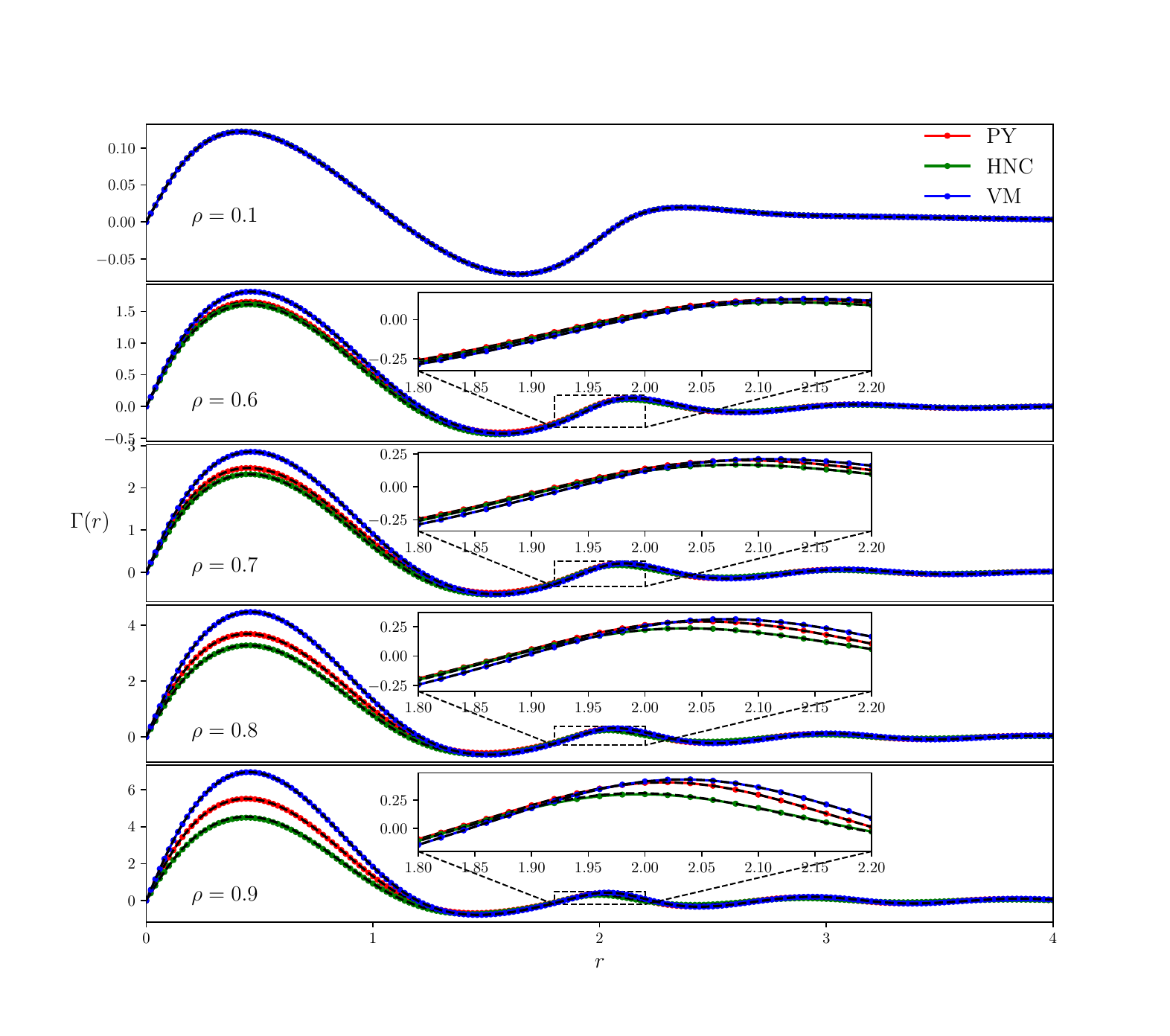}
	\caption{PINN predictions vs. simulation results of $\Gamma(r)$ for the OZ equation with the PY, HNC and VM closures. The results from the numerical simulations are depicted with black dashed lines. Remarkably, the curves from PINN predictions align so closely with the simulation results that they nearly overlap, highlighting the accuracy of the PINN approach.}
	\label{Fig_PINN_comparsion}	
\end{figure*}

In this section, the PINN approach as described in Section \ref{sec_PINN_forward} is applied to solving the OZ equation with various closures respectively. The PY, HNC, and VM closures are selected. The parameters for the OZ equation are employed as $k_B=\sigma=\epsilon=1$ and $T=2$.
The same network structure and Fourier feature embedding are employed for all the closures as well as for all the densities. As shown in Fig. \ref{Fig_PINN_comparsion}, the PINN predictions agree well with the simulation results (using a high resolution $N=2000$), highlighting the high accuracy of our PINN approach.

When $\rho=0.1$, there is almost no difference among the three closures. This is consistent with previous work\cite{Bedolla2022}. 
When the density is small, the physical system behaves like a gas, so the long-range correlation effect is very weak.
With the increase of $\rho$, the long-range correlation effect becomes stronger, as represented by the slow decay of quasi-periodic oscillation along the $r$ direction shown in Fig.\ref{Fig_PINN_comparsion}.
The prediction with VM closure shows the largest oscillation amplitude, the prediction with HNC closure has the smallest oscillation amplitude i.e., stronger long-range correlation. 
Our PINN model can capture the significant discrepancy of predictions of correlation function among closures even at higher density.

\subsection{Application to various types of fluids}
\label{sec_closures}
\begin{figure*}[htbp]
	\centering	
	\includegraphics[width=15.5cm]{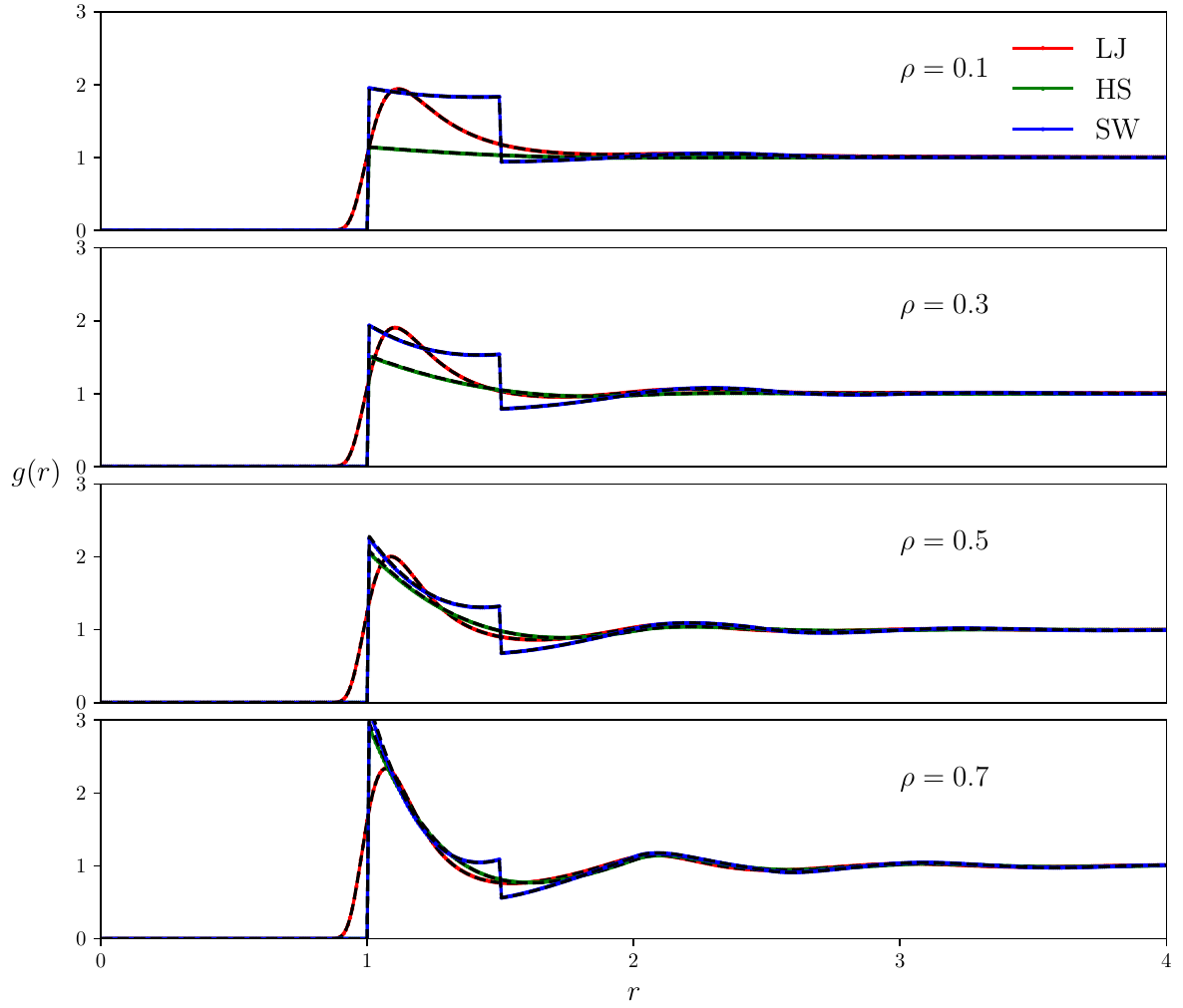}
	\caption{PINN predictions vs. simulation results of $\Gamma(r)$ for the OZ equation for the 12-6 Lennard-Jones (LJ), hard-sphere (HS), square-well (SW) fluids. The results from the numerical simulations are depicted with black dashed lines. Remarkably, the curves from PINN predictions align so closely with the simulation results that they nearly overlap, highlighting the accuracy of the PINN approach.}
	\label{Fig_PINN_comparsion_fluids}	
\end{figure*}

In the above section we presented the application of PINN to the OZ equation for the LJ fluid. In this section, the PINN approach as described in Section \ref{sec_PINN_forward} is applied to solving the OZ equation using PY closure for various fluids. To compare with the numerical results in reference\cite{bedolla2022evolutionary}, the reduced temperature is set to $T=1.5$.
The definition of fluid is represented by the potential function $u(r)$. In addition to the LJ fluid, we also consider another two widely studied fluids, namely, the hard-sphere (HS) fluid and square-well (SW) fluid,\cite{Carley1983,Santos2020} whose potential functions are defined as:
\begin{equation}
	u_{HS}=\left\{
	\begin{aligned}
		\infty, \qquad &r<\sigma \\
		0, \qquad &r\ge\sigma
	\end{aligned}
	\right.
	,
\end{equation}
\begin{equation}
	u_{SW}=\left\{
	\begin{aligned}
		\infty, \qquad &r<\sigma \\
		-\epsilon, \qquad &\sigma \le r \le \lambda \sigma \\
		0, \qquad &r>\lambda\sigma
	\end{aligned}
	\right.
	,
\end{equation}
where $\lambda=1.5$  is the interaction range and $\epsilon$ is the well depth for the square-well fluid. 
The two fluids have discontinuities in their potential functions. To better resolve the discontinuities, a fine resolution of residual points is employed with $N=4000$. 
The PINN predictions as well as the numerical simulation results are depicted in Fig. \ref{Fig_PINN_comparsion_fluids}. Note that our numerical simulation results are derived with high resolution $N=10000$, 
are consistent with previous results.\cite{Trokhymchuk2005,Henderson1976} Our PINN predictions  coincide well with the simulation results, highlighting the high accuracy of our PINN approach, particularly the excellent capability to capture the  discontinuity for the hard-sphere and square-well fluids. 
Given the significant role of thermodynamic quantities in the integral theory, we aim to further validate the accuracy of the PINN approach quantitatively. We compute the compressibility factor using the virial equation of state\cite{Tsednee2019}, especially as it has a strong connection to the discontinuous potential functions inherent to hard-sphere and square-well fluids. Its definition is given as follows:
\begin{align}
P_{HS} &= 1+\frac{2\pi}{3}\rho\sigma^3 g(\sigma^+), \\
P_{SW} &= 1+\frac{2\pi}{3}\rho\sigma^3 (g(\sigma^+)+\lambda^3(g(\lambda \sigma^+)-g(\lambda \sigma^-))),
\end{align}
where  $g(\sigma^+)$ and $g(\sigma^-)$ denote the right/left limit, respectively, of the function $g(r)$ at $r=\sigma$. Similarly, this also applies for 
 $g(\lambda \sigma^+)$ and 
$g(\lambda \sigma^-)$.
The compressibility factor $Z$ is defined as 
\begin{align}
Z = \beta{P}/\rho
\end{align}  
Table \ref{table_PINN_comparsion_fluids} lists the predictions of the compressibility factor from this work and those from the reference \cite{bedolla2022evolutionary},  showing  excellent agreement. We identified that the contribution to the total error from the discontinuity points is greater than from the middle points. And the thermodynamic property like $Z$ is mostly relevant to the values of $g(r)$ that are close to the discontinuity points. From Fig. \ref{Fig_PINN_comparsion_fluids}, we see that the short-range correlation increases with density for all the fluids, which also enhances the local density fluctuation. 
Obviously, the local fluctuation in the short range is not so strong for the potential function with a smoother attractive part such as LJ fluid. 
The prediction error of $Z$ of the hard-sphere and square-well fluids narrows with small $\rho$ such as 0.1. And the relative error of the prediction is very small ($10^{-5}$-$10^{-4}$). When the density is 0.7, the relative error of the SW fluid is still small, but increases to $10^{-2}.$ We conclude that for a high density fluid system with strong local fluctuation, more residual points are needed to reduce the error.

\begin{table*}[tbp]
	\centering
	\caption{Predictions of compressibility factor $Z$, where the reference value $Z_{ref}$ is derived from numerical simulations in the reference \cite{bedolla2022evolutionary}. }
	\begin{tabular}{ccccccc}
		\toprule
		\multirow{2}{*}{Density} & \multicolumn{3}{c}{Hard sphere fluid} & \multicolumn{3}{c}{Square well fluid} \\
		\cmidrule(r{1em}l{1em}){2-4} \cmidrule(r{1em}l{1em}){5-7}
		&$Z_{ref}$ & $Z$ &$|Z-Z_{ref}|/Z$ & $Z_{ref}$ & $Z$ &$|Z-Z_{ref}|/Z$ \\
		\midrule
		0.1 & 1.23933 & 1.23943 & 7.71E-05 & 0.77952 & 0.77967 & 1.93E-04 \\
		0.3 & 1.95377 & 1.95520 & 7.34E-04 & 0.63587 & 0.63774 & 2.93E-03 \\
		0.5 & 3.17322 & 3.17941 & 1.95E-03 & 1.11718 & 1.12576 & 7.68E-03 \\
		0.7 & 5.32284 & 5.34269 & 3.73E-03 & 3.09157 & 3.12538 & 1.09E-02 \\
		\bottomrule
	\end{tabular}
	\label{table_PINN_comparsion_fluids}
\end{table*}

\subsection{Physics-informed DeepOnet for the solution of the parameterized OZ equation}
\label{sec_deepOnet}
In subsections \ref{sec_PINN_forward_inverse} and \ref{sec_closures}, we used the PINN approach to solve the OZ equation with different closures, attaining substantial predictive accuracy. However, the PINN approach necessitates training the network for fixed values of the parameters. To cope with this restriction, we turn to the physics-informed DeepOnet (PIDeepOnet) approach to directly tackle the parameterized OZ equation. In this context, we treat the density 
$\rho \in [0.1,0.9]$  as the parameter, using the VM closure. If necessary, other thermodynamic quantities, such as temperature, can also be employed as parameters. For this case, the other parameters for the OZ equation are fixed as $k_B=\sigma=\epsilon=1$ and $T=2$.

The PDE loss function is constructed through repeated evaluations of Eq. \eqref{eq_loss_PDE}  for different values of  $\rho,$ namely
\begin{equation}\label{eq_loss_PPDE}
	\mathcal{L}_{PDE}=\frac{1}{(N_{\rho}+1)(N-1)}\sum_{j=0}^{N_{\rho}}\sum_{i=1}^{N-1}\left[{\Gamma^*(r_i; \rho_j)-\Gamma(r_i;\rho_j)}\right]^2
\end{equation}
where $\rho_j$ for $j=0,...,N_{\rho}$ are the equally-spaced points in the $\rho$ range. $N=800$ and $N_{\rho}=200$ is chosen.
The boundary condition is defined as 
\begin{equation}
	\Gamma(0; \rho_j)=\Gamma(R_{\max}; \rho_j)=0 \qquad j=0,...,N_{\rho}
\end{equation}
The loss function, a composite of the PDE loss and BC loss, is defined as
\begin{align}
	loss&=\mathcal{L}_{PDE}+\mathcal{L}_{BC}\\
	&=\mathcal{L}_{PDE} + \frac{1}{2(N_{\rho}+1)}\sum_{j=0}^{N_{\rho}}\left[\Gamma(0; \rho_j)^2+\Gamma(R_{\max}; \rho_j)^2\right]
\end{align}

\begin{figure*}[htbp]
	\centering	
	\includegraphics[width=14cm]{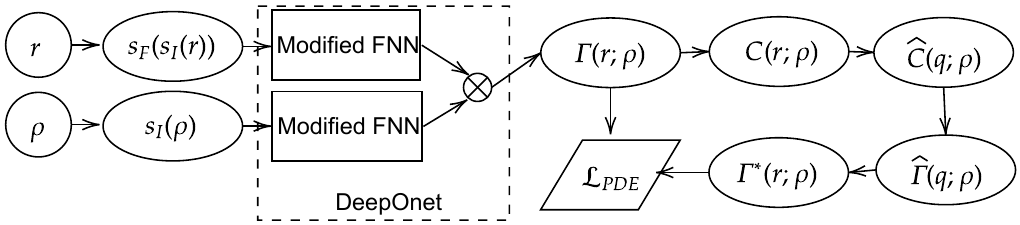}
	\caption{Forward propagation of PDE loss in the Physics-informed DeepOnet approach for solving the parameterized OZ equation. $s_I(\cdot)$ denote the normalization function of input and $s_F(\cdot)$ denotes the Fourier feature transformation function. $\otimes$ represents an inner product.}
	\label{fig_PINN_forward_PPDE}	
\end{figure*}

For the sake of clarity, the forward propagation of the PIDeepOnet  approach for deriving the PDE loss is shown in 
Fig. \ref{fig_PINN_forward_PPDE}. The steps for calculating the PDE loss are listed as follows:
\begin{enumerate}[(1)]
	\item Predict  $\Gamma(r_i; \rho_j)$ for $i=0,...,N$ and $j=0,...,N_{\rho}$
	\item Compute $C(r;\rho)$ for each point according to the HNC closure.
	\item Perform Fourier transformation for each $\rho$ according to Eq. \eqref{eq_C_Fourier}.
	\item Calculate the Fourier transformation of $\Gamma(r; \rho)$ for each $\rho$ with Eq. \eqref{eq_Gamma_OZ}.
	\item Perform inverse Fourier transformation for each $\rho$ with Eq. \eqref{eq_Gamma_inverse_Fourier} to obtain $\Gamma^*(r; \rho)$ .
	\item Calculate the PDE loss with Eq. \eqref{eq_loss_PPDE}.
\end{enumerate}

\begin{figure*}[htbp]
	\centering	
	\includegraphics[width=14cm]{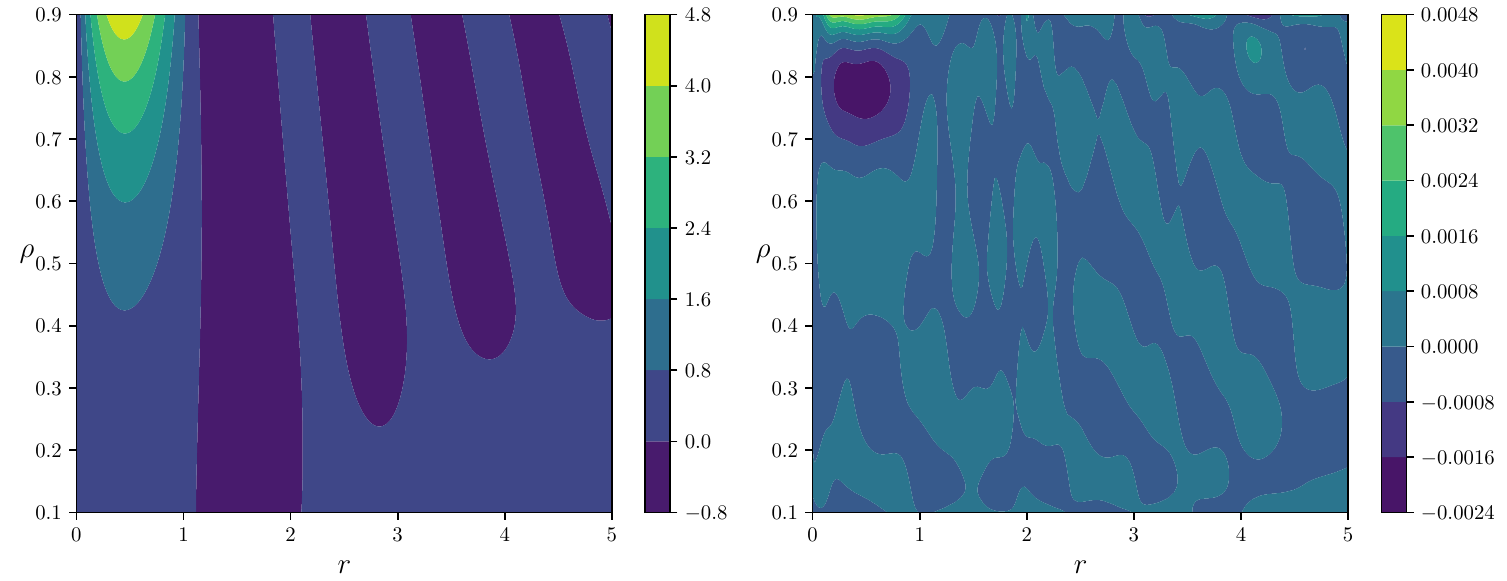}
	\caption{Predictions of $\Gamma(r)$ (left) and its absolute error (right) for parameterized OZ equations with the physics-informed DeepOnet approach. }
	\label{fig_PPDE_result}	
\end{figure*}

Both the trunk and branch neural networks use the same structure $(1, 30, 30, 30, 30, 30, 30, 1)$, each entry denotes the neuron count for that layer. Note that the trunk net takes $r$ as input and the branch net takes $\rho$ as input.
Fourier feature embedding is only applied to the trunk net with a variance $\sigma_F^2=1$ and the feature number $m=50$. Similarly, the network is pre-trained with an initialization function $y=c\rho r e^{-br}$ with $c=30$ and $b=3$. The PIDeepOnet prediction of $\Gamma(r)$ and its absolute error is shown in Fig. \ref{fig_PPDE_result}. It is evident from the contour plot of $\Gamma(r; \rho)$ that the maximum oscillation amplitude as well as the oscillation frequency increase with $\rho$. Also, the oscillation decays slower to zero with the increase of $r$, indicating the PIDeepOnet model learned accurately the relationship between the system density and the long-range correlation. The maximum absolute error of the PIDeepOnet prediction exists near the corner ($\rho=0.9, r=0$), where the absolute value of $\Gamma(r; \rho)$ is also very large. As a result, the PIDeepOnet prediction is accurate with a relative $L_2$ error $6.18 \times 10^{-4}$ over the domain $(r,\rho) \in [0, R_{\max}]\times[0.1, 0.9]$. The extremely low relative error indicates the PIDeepOnet model is an trustworthy model for the solution of the  parameterized OZ equation.

\FloatBarrier
\section{\label{sec_conclusion}Conclusion and outlook}
The OZ integral equation theory is a powerful approach for computing correlations in simple fluids. In this work, we introduce a PINN approach to solve the OZ equation and a PIDeepOnet approach for tackling the parameterized OZ equation. Both the PINN and PIDeepOnet approachs are designed with state-of-the-art techniques, 
such as Fourier feature embedding, a modified feedforward neural network, and a self-adaptive weighting training methodology, resulting in superior accuracy and efficiency. The accuracy and robustness of the PINN approach is validated by its application to both forward and inverse OZ challenges, encompassing various closures (HNC, PY and VM closures) and fluids (12-6 Lennard-Jones, hard-sphere and square-well fluids). The PIDeepOnet approach facilitates the extension to the parameterized OZ equation by  handling separately the coordinate and parameter within the DeepOnet architecture.  Its capability in accurately predicting correlation functions over various thermodynamic states, offers considerable promise for advancing thermodynamic models in liquid theory.

Our PINN approach was applied to a forward OZ problem with the HNC closure. Compared to a previous PINN approach\cite{carvalho2022physics}, our PINN approach showcases enhanced accuracy, particularly for long-range correlation at high density, coupled with a significant improvement in computational efficiency. Moreover, the PINN approach was successfully applied to an inverse OZ problem, learning a correlation function according to the experimental neutron scattering data. Furthermore, the robustness of the PINN approach was further validated through applications to forward OZ problems with various closures as well as various fluid potentials. Impressively, the PINN approach is capable to of resolving discontinuous correlation functions, however it requires a higher number of residual points. Finally, the PIDeepOnet approach is employed to solve a parameterized OZ problem, taking the density as a parameter. Significantly, the relative error is at the order of $10^{-4},$ suggesting that the PIDeepOnet methodology may serve as an effective surrogate for equation-of-state models in thermodynamics, particularly in high-dimensional contexts.

Using as an example a  homogeneous  fluid, we have demonstrated that the PIDeepOnet model is powerful. The current approach can be extended to heterogeneous fluid and multiphase fluids. Another potential area is using the current approach with the OZ equation to infer the interaction potential function as an inverse problem, thus extracting the effective pair potential in mesoscale simulation from atomic simulation or experiment data. It is also possible to generalise and extend upon the approach adopted here by learning the closure model for a wide variety of systems where the OZ framework has been applied such as the molecular OZ and RISM approaches, and for more thermodynamic properties calculation e.g., for the solvation energy of biomolecules.

\section*{Acknowledgments}
The work of WC and PG is supported by the U.S. Department of Energy, Advanced Scientific Computing Research program, under the Physics-Informed Learning Machines for Multiscale and Multiphysics Problems (PhILMs) project (Project No. 72627). The work of PS is  supported by the U.S. Department of Energy, Advanced Scientific Computing Research
program, under the Scalable, Efficient and Accelerated Causal Reasoning Operators, Graphs and Spikes for Earth and
Embedded Systems (SEA-CROGS) project (Project No. 80278).
Pacific Northwest National Laboratory (PNNL) is a multi-program national laboratory operated for the U.S. Department of Energy (DOE) by Battelle Memorial Institute under Contract No. DE-AC05-76RL01830.

\nocite{*}
\section*{\label{sec_reference}References}
\bibliography{aipsamp}

\end{document}